\documentclass[runningheads]{llncs}

\usepackage[T1]{fontenc}
\usepackage[utf8]{inputenc}
\usepackage{amsmath,amssymb,amsfonts}
\usepackage{microtype}
\usepackage{enumitem}
\usepackage{booktabs}
\usepackage{array}
\usepackage{xurl}
\usepackage{hyperref}

\hypersetup{
  colorlinks=true,
  linkcolor=blue,
  citecolor=blue,
  urlcolor=blue
}

\newcommand{\Sig}{\mathsf{Sig}}
\newcommand{\Sign}{\mathsf{Sign}}
\newcommand{\Verify}{\mathsf{Verify}}
\newcommand{\Impact}{\mathsf{Impact}}

\title{The Native-Signature Boundary in\texorpdfstring{\\}{ }
Post-Quantum Distributed Authorization}
\titlerunning{The Native-Signature Boundary in PQ Authorization}

\author{Dariia Porechna\orcidID{0009-0009-8834-4652}}
\authorrunning{D. Porechna}
\institute{EternaX Labs\\
\email{dariia.p@eternax.ai}}

\begin{document}
\raggedbottom
\maketitle

\begin{abstract}
Post-quantum signature migration poses a distinct systems problem when authorization is distributed among multiple parties.
In native threshold signing, the signature algorithm may determine key generation, share state, preprocessing, interaction, combination, refresh, and recovery.
Architectures that evaluate threshold policy outside the native signing relation can reduce this coupling, but their authorization evidence is not accepted by an unchanged native verifier unless a trusted complete-key signer translates approval into a native signature.

This paper organizes that design boundary through three properties: native-signature compatibility, unilateral-signing resistance, and threshold-layer agility.
We classify specialized threshold signatures, generic MPC signing, distributed hash-based constructions, programmable multisignature and dual-gate authorization, and threshold-authorized HSM signing.
A migration impact surface identifies which components change with the signature algorithm.
Across the surveyed families, no design simultaneously provides native output, unilateral-signing resistance, and threshold-layer agility.
This is an architectural tension, not an impossibility claim, and it clarifies why a replaceable API alone does not make distributed authorization cryptographically agile.

\end{abstract}

\section{Introduction}
\label{sec:intro}

Cryptographic agility is the capability to replace and adapt cryptographic algorithms while preserving security and ongoing operation~\cite{nist2026agility}.
That literature treats the replaceable unit as an algorithm behind a protocol, API, or infrastructure interface.
This paper specializes the question to distributed authorization, where the algorithm can also determine the authorization protocol and its persistent state.
For an ordinary signature integration, the visible work is familiar: introduce the new algorithm, register new public keys, overlap verification during migration, and retain enough metadata to validate historical signatures.
For distributed signing, the algorithm reaches further into the system.
It can determine how a joint key is generated, what each party stores, which preprocessing is required, how partial signatures are produced and combined, and how shares are refreshed after compromise.

The post-quantum transition turns this coupling into an immediate migration constraint.
ML-DSA and SLH-DSA are standardized for conventional signing, not as threshold protocols~\cite{nist2024fips204,nist2024fips205}.
Federal policy now sets agency migration deadlines and directs the development of post-quantum requirements for covered contractors~\cite{eo14412}, so custody operators must choose among distributed-signing designs whose migration properties differ substantially.
Threshold ECDSA and threshold Schnorr exploit the algebraic structure of their native signing equations~\cite{gennaro2018gg18,canetti2020cmp,komlo2020frost,rfc9591}.
Threshold lattice signatures can require different distributed sampling, rejection, abort, and proof machinery~\cite{delpino2024thresholdraccoon,delpino2025finally,celi2026efficientmldsa}.
For standardized hash-based signatures, oracle-respecting black-box threshold signing is impossible under the model of Kondi, Kumar, and Vanegas; the remaining approaches require distributed hash evaluation, preprocessing, or additional structure~\cite{kondi2026blackbox,kelsey2025distributed}.
An unchanged verification API can therefore conceal a complete replacement of the custody protocol behind it.

The opposite design is increasingly common in programmable systems.
A contract, vault, or consensus rule can verify several ordinary signatures and evaluate threshold policy separately.
Each member key can then migrate independently from ECDSA to ML-DSA or SLH-DSA.
The trade-off is that the authorization result is not one native signature accepted by an unchanged legacy account.
A trusted HSM can instead translate the same approval into native output by signing with a complete key, introducing the trust boundary analyzed in Section~\ref{subsec:hsm}.

All of these classes can plausibly be described as ``crypto-agile,'' but the label alone does not explain what actually migrates.
This paper organizes the design space around three questions:

\begin{enumerate}[leftmargin=*]
  \item Does the destination receive one signature accepted by its unchanged native verifier?
  \item Can any one participant or trusted component produce an accepted authorization?
  \item Can the signature scheme change without replacing the threshold authorization protocol and state?
\end{enumerate}

If no component may hold the complete native key, producing a native signature places the scheme's signing relation inside a specialized threshold protocol or a generic secure-computation circuit.
Conversely, keeping the threshold layer independent of the signature scheme requires either a programmable verifier or a trusted component that translates threshold approval into native signing.
The classification records this recurring architectural tension across the surveyed families.
Generic MPC is the important qualification: it can keep the execution engine stable, while the signing circuit, native key state, validation, and operational procedures still change with the signature scheme.

\paragraph{Organization.}
Section~\ref{sec:method} states the scope and method.
Section~\ref{sec:terms} defines the comparison dimensions.
Section~\ref{sec:classes} classifies the principal design families.
Section~\ref{sec:migration} applies the framework to representative ECDSA-to-SLH-DSA and ECDSA-to-ML-DSA migrations.
Section~\ref{sec:guidance} records practical consequences and limitations.

\section{Scope and method}
\label{sec:method}

This work presents a conceptual taxonomy of design choices rather than a systematic literature review, performance survey, or construction paper.
Protocol and infrastructure agility are already specified as the ability to migrate algorithms behind identifiers, APIs, and implementations~\cite{rfc7696,nist2026agility}.
NIST's Threshold Call separately requires interchangeability: the output of a threshold scheme must be usable by the corresponding conventional primitive~\cite[Sec.~3.3.2]{brandao2026thresholdcall}.
Threshold signatures have been surveyed broadly, and the classical native-threshold family for ECDSA has been surveyed in its own right~\cite{sedghighadikolaei2025survey,aumasson2020survey}.
Those sources do not share one unified framework for where scheme dependence sits when authorization is distributed and the destination verifier may or may not be programmable.
The intended contribution is that framework: comparison dimensions, the native-signature boundary, a taxonomy of representative design families, and the migration impact surface used to identify which artifacts change.

The classification is architectural rather than exhaustive.
It covers public material available through August 2026 and prioritizes standards and peer-reviewed work; preprints, workshop material, and research previews are included where mature constructions are unavailable and are identified as such.
The selected approaches expose different placements of signature-scheme dependence:

\begin{itemize}[leftmargin=*]
  \item specialized native threshold protocols for ECDSA, Schnorr, and lattice signatures;
  \item generic MPC evaluation of a native signing algorithm;
  \item distributed hash-based signatures, including context-aware threshold-PRF approaches;
  \item programmable multisignature and signature-agnostic threshold authorization;
  \item a trusted HSM or signer that releases native signatures after threshold approval.
\end{itemize}

The families classify authorization paths rather than providers: one custody offering may expose several paths.
Zodia Custody, for example, documents both an offline HSM-based custody model and an alternative MPC service~\cite{zodia2026custody}.

Here, \emph{generic MPC} means evaluating a signature scheme's signing algorithm as a circuit in a general-purpose secure-computation framework, rather than using a threshold protocol designed specifically for that signature scheme.

The comparison follows each architecture across the same lifecycle:
key generation and setup, persistent state, approval, combination, verification, maintenance, policy, evidence, and integration.
For each family we ask which lifecycle components depend on the signature scheme, whether a complete native key exists, and what the destination verifies.

The scope excludes performance rankings.
Protocol cost depends on parameters, adversary model, network, implementation, and hardware.
The classification entries characterize each intended authorization path under its configured policy and assume that every component enforces its specified rules; availability, administrative override paths, and implementation quality remain separate deployment concerns.
We also do not claim that one architecture dominates.
A fixed-verifier system may rationally prefer scheme-coupled native threshold signing, while a programmable asset-control system may not need a native signature at all.

\section{Comparison dimensions}
\label{sec:terms}

Let $\Sig=(\mathsf{KeyGen},\Sign,\Verify)$ be a signature scheme and let $M$ be the operation to authorize.

\begin{definition}[Native-signature compatibility]
\label{def:native}
An authorization is \emph{native for $\Sig$} if its complete output is one signature $\sigma$ accepted exactly by
\[
  \Verify(\mathsf{pk},M,\sigma)=1,
\]
without auxiliary proofs, threshold-specific verification state, or a modified verification rule.
\end{definition}

This definition specializes the NIST Threshold Call's interchangeability criterion: the output of a threshold scheme must be usable by a subsequent operation of the corresponding conventional primitive~\cite[Sec.~3.3.2]{brandao2026thresholdcall}.
A contract that counts $t$ ordinary signatures may be application-compatible, but it is not native for an account that previously accepted one ECDSA signature.
The strict definition matters in systems such as Bitcoin, where spending a signature-locked output ultimately requires a signature accepted by its designated ECDSA or Schnorr verification path.

\begin{definition}[Unilateral-signing resistance]
\label{def:custody}
For $2\leq t\leq n$, an architecture provides \emph{$t$-of-$n$ unilateral-signing resistance} if, throughout setup and operation, no participant, dealer, or trusted component possesses or can reconstruct material sufficient to authorize alone, and an adversary controlling fewer than $t$ authorization parties cannot produce an accepted authorization.
\end{definition}

This criterion concerns the compromise boundary rather than the output format or sharing mechanism.
A programmable quorum can satisfy it without sharing one native key.
An HSM holding the complete key does not satisfy it because compromise of its key-use boundary creates a unilateral signer, even when normal operation requires $t$ signed approvals.

\begin{definition}[Threshold-layer agility]
\label{def:agility}
An architecture has \emph{threshold-layer agility} across a family of signature schemes if replacing one scheme with another does not replace its threshold relation, threshold-specific protocol, or threshold secret state.
Distributed setup, approval and combination semantics, maintenance, and threshold-policy evaluation remain unchanged.
Independently held member keys, ordinary signing and verification adapters, algorithm-acceptance rules, and evidence encodings may change; when the threshold relation only gates a trusted signer, that signer's native key and signing implementation may also change.
Secret-shared native signing material counts as threshold secret state.
\end{definition}

The boundary is functional: secret shares used to compute the native signing relation lie inside the threshold layer, whereas a complete-key signer's implementation lies in a downstream translation component gated by threshold approval.
This is intentionally stronger than format, API, or verifier agility.
An algorithm identifier may select a new threshold protocol while every participant replaces shares and preprocessing.
A stable custody API may hide those changes from application code.
While both are useful engineering properties, neither makes the threshold layer scheme-independent.

\subsection{Migration impact surface}
\label{subsec:blast}

For an architecture $\mathcal A$, let
\[
\begin{aligned}
\mathcal C_{\mathcal A}=\{&
\mathsf{keygen},\mathsf{setup},\mathsf{state},\mathsf{approve},
\mathsf{combine},\\[-2pt]
&\mathsf{verify},\mathsf{maintain},\mathsf{policy},
\mathsf{evidence},\mathsf{integration}\}.
\end{aligned}
\]

Here, $\mathsf{keygen}$ and $\mathsf{setup}$ cover creation and installation of signing material; $\mathsf{state}$ covers persistent authorization data; $\mathsf{approve}$ and $\mathsf{combine}$ cover online contribution and aggregation semantics; $\mathsf{verify}$ covers member and destination verification; $\mathsf{maintain}$ covers refresh, backup, and recovery; $\mathsf{policy}$ covers quorum and algorithm acceptance; $\mathsf{evidence}$ covers retained authorization artifacts; and $\mathsf{integration}$ covers encodings and system interfaces.
In $\Sig_0\rightarrow\Sig_1$, $\Sig$ denotes the accepted native-output scheme for native architectures and the independently verified member-signature scheme for programmable architectures.

The \emph{migration impact surface}
\[
  \Impact_{\mathcal A}(\Sig_0\rightarrow\Sig_1)
  \subseteq \mathcal C_{\mathcal A}
\]
is the set of components whose implementation, persistent state, wire semantics, trust assumption, failure procedure, or security argument must change with the scheme.
It is end-to-end: destination-verifier and interface changes are included rather than treated as external prerequisites.
The set is more informative than its size.
Replacing a stateless verifier library is not operationally equivalent to rerunning a distributed-key ceremony or invalidating presignature state.

\subsection{The native-signature boundary}
\label{subsec:boundary}

Native verification fixes a scheme-specific relation
\[
  R_{\Sig}=\{(\mathsf{pk},M,\sigma):
  \Verify(\mathsf{pk},M,\sigma)=1\}.
\]
Some component must produce an output in $R_{\Sig}$ from native signing material.
There are three broad placements:

\begin{enumerate}[leftmargin=*]
  \item a specialized threshold protocol jointly realizes the signing relation;
  \item a generic MPC engine securely evaluates a circuit for $\Sign$;
  \item one component holds or reconstructs the complete native key and signs after checking distributed approval.
\end{enumerate}

If none of these is acceptable, the destination must verify a different authorization relation.
The enumeration serves as a bookkeeping rule: generic MPC localizes scheme dependence to a circuit, specialized threshold signing builds it into a protocol, programmable authorization moves it into the verifier, and a trusted signer concentrates it in the complete-key component.

\section{Classification of authorization designs}
\label{sec:classes}

\begin{table}[t]
\caption{Classification under Definitions~\ref{def:native}--\ref{def:agility}. ``Agile'' refers to the threshold authorization layer, not to every scheme-dependent component.}
\label{tab:classification}
\centering
\scriptsize
\setlength{\tabcolsep}{2.5pt}
\begin{tabular}{@{}>{\raggedright\arraybackslash}p{2.65cm}
>{\centering\arraybackslash}p{1.05cm}
>{\centering\arraybackslash}p{1.70cm}
>{\centering\arraybackslash}p{1.55cm}
>{\raggedright\arraybackslash}p{4.00cm}@{}}
\toprule
\textbf{Design family} & \textbf{Native}\newline\textbf{output} &
\textbf{No unilateral}\newline\textbf{signer} &
\textbf{Threshold-layer}\newline\textbf{agility} &
\textbf{Scheme dependence or trust placement} \\
\midrule
Specialized threshold signatures & Yes & Yes\textsuperscript{*} & No &
Distributed protocol, key shares, and preprocessing \\
\addlinespace
Generic MPC of $\Sign$ & Yes & Yes\textsuperscript{*} & No &
Signing circuit and native key shares; MPC engine may remain \\
\addlinespace
Distributed hash-based signatures & Yes & Conditional & No &
Distributed evaluation of the hash-based signing relation \\
\addlinespace
Programmable multisignature & No & Yes & Yes &
Ordinary member-signature verification adapter \\
\addlinespace
Dual-gate authorization & No & Yes & Yes &
Member-signature adapter; threshold authorization state remains \\
\addlinespace
Threshold-authorized HSM & Yes & No & Yes &
Complete native key and scheme implementation in a trusted signer \\
\bottomrule
\end{tabular}
\vspace{0.4ex}

\parbox{\linewidth}{\textsuperscript{*} Requires dealerless distributed key generation and a malicious-secure signing protocol or MPC backend.}
\end{table}

Table~\ref{tab:classification} applies the three definitions.
The hash-based entry is construction-dependent: the cited stateful construction uses a trusted dealer, while the PRAWNS preview leaves distributed key generation open.

\subsection{Interpreting the empty cell}

Across the surveyed families, Table~\ref{tab:classification} has no row that is native, has unilateral-signing resistance, and has threshold-layer agility under Definitions~\ref{def:native}--\ref{def:agility}.
This illustrates a recurring design tension within the classification; it does not assert a general impossibility theorem.

Native output with unilateral-signing resistance requires the signing relation to be evaluated without placing the complete key in one component.
Specialized threshold signatures embed that relation in a distributed protocol.
Generic MPC instead embeds it in a circuit evaluated over shared native key material.
In both cases, changing the signature scheme changes artifacts inside the distributed signing system.
Moving scheme dependence out of that system requires either a programmable verifier, which accepts a non-native authorization relation, or a trusted signer, which restores native output by holding the complete key.

Generic MPC is the closest qualification because its execution engine and protocol backend may remain stable.
The replacement circuit must nevertheless implement the new signing behavior, the native key shares must be replaced, and scheme-specific validation, preprocessing, failure handling, and security arguments may require review.
This is engine agility rather than the threshold-layer agility of Definition~\ref{def:agility}.

The remainder of this section applies a common profile to each family: native output, unilateral-signing resistance, threshold-layer agility, and the locus of migration.
The accompanying discussion explains the qualifications hidden by the compact entries.

\subsection{Specialized native threshold signatures}
\label{subsec:specialized}

\begin{itemize}[leftmargin=*,nosep]
  \item \textbf{Native output:} Yes
  \item \textbf{Unilateral-signing resistance:} Yes, with dealerless setup
  \item \textbf{Threshold-layer agility:} No
  \item \textbf{Migration locus:} Distributed protocol, native key shares, and any scheme-specific preprocessing, refresh, or security argument
\end{itemize}

Threshold ECDSA, FROST, and threshold lattice signatures produce a signature in the same format accepted by the same unmodified verifier as their single-signer counterparts~\cite{gennaro2018gg18,rfc9591,delpino2024thresholdraccoon,celi2026efficientmldsa}.
The destination runs an unchanged native verifier, and no participant needs to hold the complete key during signing.
NIST's notes on threshold EdDSA/Schnorr state this property: such threshold signatures can serve as a drop-in replacement for conventionally produced signatures without changing legacy verification code~\cite{brandao2022eddsa}.
Custody-infrastructure providers describe deployed instances of this pattern: Fireblocks applies MPC-CMP to blockchain ECDSA and EdDSA signatures without gathering the complete private key, while BitGo's MPC/TSS wallets produce one standard signature that appears on chain as a single-key transaction~\cite{fireblocks2026mpccmp,bitgo2026tss}.
The classical ECDSA branch is the most developed instance and has been surveyed in its own right~\cite{aumasson2020survey}.
This family-level classification does not erase construction-specific setup assumptions: Threshold Raccoon, for example, uses trusted centralized key generation and therefore does not satisfy Definition~\ref{def:custody} across the complete lifecycle~\cite{delpino2024thresholdraccoon}.

The distributed protocol is nevertheless scheme-specific.
Its key-generation or sharing setup follows the native key structure, and its online phase follows the signing equation.
When the construction provides preprocessing or share refresh, those mechanisms must follow the scheme's randomness, abort behavior, and joint-key structure.
Moving between unrelated signature families therefore changes most of the signing lifecycle even if the external API remains \texttt{sign(message)}.

\subsection{Generic MPC signing}
\label{subsec:generic}

\begin{itemize}[leftmargin=*,nosep]
  \item \textbf{Native output:} Yes
  \item \textbf{Unilateral-signing resistance:} Yes, with dealerless setup and a suitable malicious-secure backend
  \item \textbf{Threshold-layer agility:} No
  \item \textbf{Migration locus:} Signing circuit, native key shares, validation, and potentially preprocessing; the MPC engine may remain
\end{itemize}

General secure computation can in principle evaluate any efficiently computable signing algorithm on shared key material by expressing it as a circuit and supplying distributed randomness and state as required~\cite{goldreich1987gmw,sedghighadikolaei2025survey}.
This completeness result does not itself provide malicious security, robust distributed key generation, unbiased randomness, fairness, or output-distribution correctness; those properties depend on the selected MPC protocol and system construction.
This includes lattice- and hash-based signing algorithms in the computability sense, but not necessarily in a practically useful form.
Hash evaluation can be circuitized, but evaluating standardized hashes inside MPC can be prohibitive~\cite{kondi2026blackbox}.
Lattice arithmetic, sampling, and rejection logic can likewise be expressed for MPC, but implementations must account for protocol cost, abort behavior, and the resulting signature distribution~\cite{bienstock2026quorus,celi2026efficientmldsa}.
This supplies native output and distributed custody without requiring a purpose-built threshold protocol for every scheme.
The MPC engine may remain stable across algorithms, and general-purpose frameworks make that reuse concrete: one high-level program can be compiled against many protocol backends and security models~\cite{keller2020mpspdz}.

The circuit is still a migration artifact.
The distributed implementation must reproduce the standard's key generation and validation requirements, while the signing circuit must reproduce its output distribution, rejection behavior, and fault handling.
Native key shares change with the scheme, and preprocessing and cost can change with the circuit.
Generic MPC is therefore engine-agile, but it does not satisfy Definition~\ref{def:agility} when migration replaces native key shares or scheme-specific preprocessing.

\subsection{Distributed hash-based signatures}
\label{subsec:hash}

\begin{itemize}[leftmargin=*,nosep]
  \item \textbf{Native output:} Yes
  \item \textbf{Unilateral-signing resistance:} Construction-dependent
  \item \textbf{Threshold-layer agility:} No
  \item \textbf{Migration locus:} Distributed evaluation of the hash-based signing relation, including construction-specific setup or reference state
\end{itemize}

SLH-DSA derives FORS and WOTS+ secret values from a secret seed using SHA-2- or SHAKE-based functions and authenticates them through a hypertree~\cite{nist2024fips205}.
Linear shares of the seed do not produce linear shares of those pseudorandom outputs.
Kondi, Kumar, and Vanegas capture the resulting obstacle in an oracle-respecting model with no dealer or preprocessing: a protocol secure against a malicious majority must distributively evaluate the hash function~\cite{kondi2026blackbox}.
Kelsey, Lang, and Lucks instead give efficient transformations for stateful LMS- and XMSS-like signatures using a trusted dealer and large common reference value; their construction does not instantiate stateless SLH-DSA~\cite{kelsey2025distributed}.

The PRAWNS research preview proposes another placement.
It uses a context-aware threshold PRF to release message-selected Winternitz secret values while preventing contributions for different messages from being pooled~\cite{prawns2026}.
The preview claims an ordinary hash-based signature that hides the threshold and participant set, subject to $t>(n+f)/2$ for $f$ corrupt parties; it explicitly leaves distributed key generation unresolved.
The proposed design aims for native, distributed signing, but its protocol is tied to the structure of the hash-based signing relation.
Changing to ML-DSA, Falcon, or another unrelated family still changes the distributed computation.

\subsection{Programmable authorization}
\label{subsec:programmable}

A programmable verifier can count ordinary member signatures or evaluate threshold authorization separately from authentication.
Member keys are independent and can migrate one at a time.
Mixed-scheme quorums support staged migration, and the threshold policy need not change with the member-signature algorithm.

The word multisignature spans both sides of the boundary, so we separate the two senses.
Key-aggregating multisignatures such as MuSig2 output one ordinary Schnorr signature under an aggregate public key~\cite{nick2021musig2}; by Definition~\ref{def:native} they are native.
MuSig2 is not generally a threshold-signature protocol, but it lies on the native side of the boundary because its participants jointly realize the native Schnorr signing relation.
The programmable sense is different: the verifier checks several ordinary signatures and applies a counting rule.

\paragraph{Multisignature.}
\begin{itemize}[leftmargin=*,nosep]
  \item \textbf{Native output:} No
  \item \textbf{Unilateral-signing resistance:} Yes
  \item \textbf{Threshold-layer agility:} Yes
  \item \textbf{Migration locus:} Independently held member keys, ordinary verification adapter, and algorithm-acceptance policy
\end{itemize}

The simplest programmable instance, $t$-of-$n$ multisignature, is deployed as a contract-validated counting rule.
Safe (formerly Gnosis Safe) smart accounts are a widely used example: an account accepts a configured threshold of valid owner authorizations over a transaction and does not produce one native account signature~\cite{safe2024}.
With $t\geq 2$, the core owner-authorization path resists unilateral authorization and provides attribution when no enabled module or administrative path can bypass the quorum; it uses no shared signing material and provides no share refresh.

\paragraph{Dual-gate authorization.}
\begin{itemize}[leftmargin=*,nosep]
  \item \textbf{Native output:} No
  \item \textbf{Unilateral-signing resistance:} Yes
  \item \textbf{Threshold-layer agility:} Yes
  \item \textbf{Migration locus:} Member keys and ordinary verification adapter; threshold authorization state and refresh remain
\end{itemize}

The dual-gate architecture adds a Shamir-shared authorization seal independent of the member-signature scheme~\cite{porechna2026dualgate}.
It preserves threshold secret state and refresh while making member-scheme replacement a key rotation.

Both approaches require an enforcement layer capable of checking their authorization relation, such as a smart account~\cite{safe2024,erc4337}, vault, consensus rule, or issuer module.
Their output is not one signature accepted by an unchanged legacy verifier.

\subsection{Threshold-authorized HSM signing}
\label{subsec:hsm}

\begin{itemize}[leftmargin=*,nosep]
  \item \textbf{Native output:} Yes
  \item \textbf{Unilateral-signing resistance:} No
  \item \textbf{Threshold-layer agility:} Yes
  \item \textbf{Migration locus:} Complete native key and signing implementation in the trusted component; the approval gate may remain
\end{itemize}

An HSM or signer service may hold the complete native key and release a signature only after verifying threshold approval.
One possible custody pattern uses a rule engine, inside or adjacent to the HSM, to check ordinary approver signatures over the intent before signing under a master key~\cite{taurus2026quantum}.
The approval layer can be signature-agnostic, while the destination receives a signature accepted by an unmodified native verifier.

This is often a practical migration bridge.
It provides distributed approval around a hardware-held key, but not resistance to unilateral authorization under Definition~\ref{def:custody}.
The HSM's key-use boundary is therefore a logical single compromise point: practical API-level attacks against HSMs are known, and a compromise that permits unauthorized use of the complete key bypasses the external quorum~\cite{focardi2021hsm}.
Enforcing approval policy inside the same trusted boundary can protect against compromise of external policy code, but it does not provide unilateral-signing resistance under Definition~\ref{def:custody}.

\section{Migration comparison}
\label{sec:migration}

Consider a custody system moving from threshold ECDSA to SLH-DSA.
The destination must first support SLH-DSA verification in every native-output architecture.
Separately, the custody system must become capable of producing the new artifact.
Adding an opcode, precompile, or verification library addresses the first event, not the second.
No cited purpose-built construction currently provides a complete dealerless threshold instantiation of SLH-DSA; the corresponding row in Table~\ref{tab:migration} is therefore a counterfactual impact profile, not a deployment claim.

\begin{table}[t]
\caption{Illustrative end-to-end migration from threshold ECDSA to SLH-DSA. Exact impact depends on the implementation; the purpose-built row is hypothetical.}
\label{tab:migration}
\centering
\small
\setlength{\tabcolsep}{3.3pt}
\begin{tabular}{@{}>{\raggedright\arraybackslash}p{2.45cm}
>{\raggedright\arraybackslash}p{4.0cm}
>{\raggedright\arraybackslash}p{3.45cm}
>{\raggedright\arraybackslash}p{1.6cm}@{}}
\toprule
\textbf{Architecture} & \textbf{Illustrative migration impact} &
\textbf{What the destination verifies} & \textbf{Native signing material} \\
\midrule
Purpose-built distributed SLH-DSA (if constructed) &
\(\mathsf{keygen}\), \(\mathsf{setup}\), \(\mathsf{state}\), \(\mathsf{approve}\), \(\mathsf{combine}\), \(\mathsf{verify}\), \(\mathsf{maintain}\), \(\mathsf{policy}\), \(\mathsf{evidence}\), and \(\mathsf{integration}\): DKG, shares, signing, combination, refresh, proofs, and destination support &
One native SLH-DSA signature &
Secret-shared \\
\addlinespace
Generic MPC of $\Sign$ &
\(\mathsf{keygen}\), \(\mathsf{setup}\), \(\mathsf{state}\), \(\mathsf{approve}\), \(\mathsf{verify}\), \(\mathsf{maintain}\), \(\mathsf{policy}\), \(\mathsf{evidence}\), and \(\mathsf{integration}\): circuit, native key shares, validation, cost profile, and destination support; engine may remain &
One native SLH-DSA signature &
Secret-shared \\
\addlinespace
Programmable multisig or dual-gate &
\(\mathsf{keygen}\), \(\mathsf{state}\), \(\mathsf{verify}\), \(\mathsf{policy}\), \(\mathsf{evidence}\), and \(\mathsf{integration}\): member keys, registry, verification adapter, and evidence encoding; threshold setup, approval semantics, combination, and maintenance remain &
Several signatures or non-native threshold evidence &
None \\
\addlinespace
Threshold-authorized HSM &
\(\mathsf{keygen}\), \(\mathsf{setup}\), \(\mathsf{state}\), \(\mathsf{verify}\), \(\mathsf{maintain}\), \(\mathsf{policy}\), \(\mathsf{evidence}\), and \(\mathsf{integration}\): HSM native key, signing implementation, and destination support; approval gate may remain &
One native SLH-DSA signature &
Complete key in HSM \\
\bottomrule
\end{tabular}
\end{table}

Table~\ref{tab:migration} instantiates the impact surface at component level and records the artifacts responsible for membership.
The sets are not universal measurements, but their end-to-end scope consistently includes destination verification, algorithm-acceptance policy, retained evidence, and integration.
They show why an external API is an incomplete agility measure.
The hypothetical purpose-built and generic MPC rows may expose the same API and produce signatures accepted by the same native verifier, while requiring materially different internal migration work.
In the programmable row the threshold layer is not among the components that change; the destination no longer verifies one native signature.
The HSM row keeps both a stable approval gate and native output by accepting a complete-key trust anchor.

The same comparison applies to a move from threshold ECDSA to ML-DSA~\cite{nist2024fips204,celi2026efficientmldsa}.
Destination support remains separate from the ability to produce the new artifact.
What changes internally is different: a specialized lattice threshold protocol exists, generic MPC evaluates a different circuit, and the hash-based distributed constructions of Section~\ref{subsec:hash} are no longer the native family.
Programmable and dual-gate rows still change member keys and adapters; the HSM row still replaces the hardware-held key and its signing implementation.
Among the surveyed approaches, SLH-DSA illustrates the more extensive distributed-evaluation path, while ML-DSA has specialized lattice threshold protocols with a different migration impact.

\subsection{Hybrid operation and historical evidence}
\label{subsec:hybrid}

Hybrid migration also depends on the layer.
A composite native signature defines a new native signing and verification relation, and RFC 9955 classifies the properties such a construction can be designed to provide, including proof composability, non-separability of the component signatures, backwards and forwards compatibility, and simultaneous verification~\cite{rfc9955}.
A signature-agnostic authorization layer can instead require a policy-selected subset of classical and post-quantum member signatures without changing its threshold state.

The two are not interchangeable in the properties they deliver.
Those RFC 9955 properties are stated for a single hybrid signature scheme, whereas a policy-selected quorum verifies its member signatures independently: the components remain separable, verification is not simultaneous, and security is bounded by the weakest quorum permitted by policy unless the policy explicitly requires a post-quantum subset.
The trade-off runs through the whole classification: the quorum keeps hybrid selection out of the signing relation, at the cost of non-separability or simultaneous-verification guarantees that a composed construction may explicitly provide.

Historical verification remains part of the migration impact surface.
Native architectures must retain old verifiers or checkpoint past validity.
Programmable architectures must retain policy versions, member registries, algorithm identifiers, and threshold-state commitments.
An architecture is not operationally agile if activation of a new primitive makes old authorization evidence uninterpretable.

\section{Practical interpretations}
\label{sec:guidance}

\subsection{Separation of claims}
\label{subsec:claims}

A system may accurately claim any of the following without satisfying the others:

\begin{itemize}[leftmargin=*]
  \item its format or API can select several signature algorithms;
  \item its verifier can overlap and retire algorithms;
  \item its MPC engine can evaluate several scheme circuits;
  \item its member-signature scheme can change without replacing threshold state;
  \item it produces a native signature from a distributed key;
  \item it requires distributed approval before a concentrated key can sign.
\end{itemize}

Therefore the ``is it crypto-agile?'' does not answer the useful question, but ``which unit can be replaced independently?'' does.

\subsection{Locating a deployment in the design space}
\label{subsec:choice}

The framework does not rank the families.
It records which of them a given constraint leaves available, and what the remaining options cost:

\begin{itemize}[leftmargin=*]
  \item A destination that accepts only one native signature excludes programmable authorization. If no complete key may exist, the distributed protocol or signing circuit becomes part of migration; otherwise, a trusted signer can preserve the approval gate while replacing its native key and implementation.
  \item A programmable destination admits every family, so the open comparison is between importing a scheme-specific threshold protocol and confining migration impact to member keys, registries and adapters, acceptance policy, and evidence encoding.
  \item Admitting a certified HSM as a complete-key holder provides a stable approval gate together with native output, in exchange for a trust anchor that has to be stated rather than assumed.
  \item Selecting generic MPC moves the review onto the scheme circuit, native key generation, distributional correctness, and cost profile, since those are the artifacts that change with the algorithm.
\end{itemize}

\subsection{Limitations}
\label{subsec:limitations}

The taxonomy is definition-sensitive.
An operator may reasonably call a generic MPC engine agile because only its loaded circuit changes.
Our threshold-layer definition is stricter because circuit replacement, native key replacement, and scheme-specific validation are material custody events.
Naming the layer resolves more disagreement than choosing one universal definition.

The paper does not benchmark the classified systems or provide a universal cost ordering.
Nor does the empty cell in Table~\ref{tab:classification} establish that no construction can satisfy all three definitions; it records the placement of scheme dependence in the representative families surveyed here.
The SLH-DSA purpose-built migration profile is hypothetical.
The paper discusses PRAWNS from an initial public research description rather than a complete paper and proof; that preview leaves distributed key generation unresolved and imposes a quorum-intersection bound.
Finally, signature agility is only one dependency layer: hashes, commitments, secure channels, HSM firmware, consensus authentication, and recovery credentials may have independent migration paths.

\section{Conclusion}
\label{sec:conclusion}

Post-quantum verification support and post-quantum distributed authorization are separate migration problems.
Native threshold signatures preserve an unchanged verifier by bringing the native signing relation into the distributed protocol.
Generic MPC places that relation in a circuit.
Programmable authorization keeps the threshold layer independent by asking the verifier to accept a different artifact.
Threshold-authorized HSMs translate agile approval into native signing while concentrating the native key in hardware.

None of these placements is universally preferable.
Among the surveyed families, preserving both native output and unilateral-signing resistance places scheme-dependent artifacts inside distributed signing; moving those artifacts outside the threshold layer requires either a programmable verifier or a complete-key trust anchor.
The classification supplies narrower language for comparing them:
what does the destination verify, where can the complete native key exist, which threshold state depends on the scheme, and what must move during migration?
Answering those questions makes an agility claim testable.

\bibliographystyle{splncs04}
\bibliography{references}

\end{document}